\documentclass[aps,prb,reprint]{revtex4-2}
\usepackage{physics} 
\usepackage{siunitx} 
\usepackage{enumerate} 
\usepackage{pgfplots}
\usepackage{pgfplotstable}
\usepackage{tikz,pgfplots}
\usepackage{pifont}
\usepackage{braket}
\usepackage{bm}
\usepackage{amssymb}
\usepackage{bbm}
\usepackage{amsmath}
\usepackage{booktabs}  
\usepackage{amsthm}
\usepackage{tabularx}
\usepackage{yhmath}
\usepackage{float}
\usepackage[graphicx]{realboxes}
\usepackage{hyperref}
\usepackage{my_symbols}
\begin{document}

\title{Anomalies in mirror symmetry enriched topological orders}

\author{Zhaoyang Ding}
\affiliation{State Key Laboratory of Surface Physics and Department of Physics, Fudan University, Shanghai 200433, China}

\author{Yang Qi}
\thanks{qiyang@fudan.edu.cn}
\affiliation{State Key Laboratory of Surface Physics and Department of Physics, Fudan University, Shanghai 200433, China}

\date{\today}
\begin{abstract}
  Two-dimensional mirror symmetry enriched topological (SET) orders can be studied using the folding approach: it can be folded along the mirror axis and turned into a bilayer system on which the mirror symmetry acts as a $\mathbb Z_2$ layer-exchange symmetry.
  How mirror symmetry enriches the topological order is then encoded at the mirror axis, which is a gapped boundary of the folded bilayer system.
  Based on anyon-condensation theory, we classify possible $\mathbb Z_2$-symmetric gapped boundaries of the folded system.
  In particular, we derive an $H^2$ obstruction function, which corresponds to an $H^3$ obstruction for topological orders enriched by the time-reversal symmetry instead of mirror symmetry.
  We demonstrate that states with a nontrivial $H^2$ obstruction function can be constructed on the surface of a three-dimensional mirror SET order.
\end{abstract}

\maketitle
\section{Introduction}
\label{sec:intro}

Topological orders~\cite{Wen1990,Wen2015nsr} are quantum many-body phases containing fractional anyon excitations, which are beyond Landau's paradigm of classifying phases by symmetry breaking.
Topological orders can be further enriched by global symmetries, forming symmetry-enriched topological (SET) orders~\cite{XGWenPSG2002,Kitaev2006,XieChen2010LUT,Mesaros2013}.
In SET orders, anyons can be transmuted by symmetry actions, and they may carry nontrivial symmetry fractionalization~\cite{Essin2013,YQi2015,Tarantino_2016,Heinrich2016,MengCheng2017}.
These nontrivial symmetry transformations of anyons not only provide experimental signatures for the detection of SET phases such as quantum spin liquids, but they can also create nontrivial quantum degrees of freedom at symmetry defects, which are useful for the storage and manipulation of quantum information.
Therefore, the classification of possible SET phases is an important problem in the study of topological phases.

For on-site symmetries, the SET phases can be classified using the $G$-crossed category theory~\cite{barkeshli2019symmetry}.
However, for crystalline symmetries, including the mirror symmetry, the classification of SET phases has not been completely solved.
The crystalline equivalence principle proposed by \citet{ryan2018cep} asserts that classification of crystalline SET phases can be obtained by treating the crystalline symmetries as on-site symmetries with the same group structure.
For example, the classification of mirror SET states is isomorphic to the classification of time-reversal SET states, which has been studied extensively~\cite{wang2017anomaly,Tachikawa2017,Tachikawa2017b,barkeshli2018time,Cordova2018, Barkeshli2020Reflection}.
However, this equivalence principle only gives an isomorphism between the two classification problems, without the details of crystalline SET phases such as the exact form of symmetry action on anyons and how to construct such SET phases.
Reference~\cite{qi2019folding} proposed that mirror-SET phases can be understood by considering the anyon-condensation boundary condition on the mirror axis, using a folding approach.
However, previous works~\cite{qi2019folding,mao2020mirror} use the tunneling matrix~\cite{lan2015gapped} to describe the anyon condensation, which does not fully characterize the boundary condition in the most general cases.
In particular, it is not clear how an analogy of the $H^3$ obstruction for onsite symmetries is represented in this formulation.

In this work, we present a detailed study of mirror SET phases, using a complete set of consistency conditions for a symmetric boundary condition on the mirror axis.
Following the folding approach in Ref.~\cite{qi2019folding}, the mirror SET can be fully characterized by a boundary condition specifying how the two regions divided by the mirror axis are connected at the axis.
We can then fold the two-dimensional (2D) system along the mirror axis, and turn the whole plane into a bilayer half-plane region, where the mirror axis becomes its boundary.
The mirror symmetry relating the two sides of the mirror axis now becomes an onsite symmetry exchanging the two layers.
In particular, by relabeling the anyons on one layer, we can treat arbitrary mirror SET states with a canonical choice of the bilayer system that depends only on the underlying topological order.
The mirror enrichment of the topological order is then fully encoded in the anyon-condensation boundary condition on the mirror axis.
We then describe this $\mathbb Z_2$-symmetric boundary with data representing how anyons condense on the boundary and how the symmetry acts on the anyon-condensation spaces, including the vertex lifting coefficients (VLCs) and boundary $U$ symbols.
Using a general theory of symmetric anyon-condensation boundary conditions, we derive a full set of consistency conditions on these data, and formulas for computing two obstruction classes characterize whether the mirror SET is anomalous.
Consistent with the crystalline equivalence principle, we provide a correspondence between the VLCs and boundary $U$ symbols describing a mirror SET, and the bulk $U$ and $\eta$ symbols describing a time-reversal SET.

The rest of the paper is organized as follows:
In Sec.~\ref{sec:bdry-set}, we briefly review the classification of a symmetric gapped boundary of SET orders,
and we introduce the VLCs and boundary $U$ symbols.
In particular, we derive a set of consistency equations these symbols must satisfy, and a formula for an obstruction class $\mathcal O$ that must vanish when solutions to the consistency equations exist.
This obstruction class, therefore, indicates whether the boundary is anomalous.
Because this obstruction class belongs to a second-cohomology group, we shall refer to it as a boundary $H^2$ anomaly.
In Sec.~\ref{thi}, we discuss how this boundary $H^2$ anomaly can be resolved by replacing the vacuum with another SET on the other side of the boundary.
In Sec.~\ref{for}, we apply the general theory in Sec.~\ref{sec:bdry-set} to study mirror SET.
Using the folding approach, we argue that a mirror SET can be fully characterized using boundary $U$ symbols on the boundary of the folded bilayer systems.
In Sec.~\ref{fif}, we focus on the boundary $H^2$ anomaly of the folded bilayer system.
We derive a formula for explicit calculation of this anomaly, and we show that it can be resolved by putting the mirror SET on the boundary of a three-dimensional (3D) SET state, using the resolution in Sec.~\ref{thi}.
In Sec.~\ref{sec:correspondence}, we provide a correspondence between the VLCs and boundary $U$ symbols characterizing the mirror SET, and the bulk $U$ and $\eta$ symbols characterizing a time-reversal SET, including the correspondence between the boundary $H^2$ and $H^3$ anomalies for the former, and the $H^3$ and $H^4$ anomalies for the latter.
This verifies the crystalline-equivalence principle~\cite{ryan2018cep} between mirror and time-reversal symmetries.
In Sec.~\ref{sec:d16}, we discuss the example of an anomalous mirror SET based on the $\mathbb{D}_{16}$ gauge theory.
This example is studied in Ref.~\cite{qi2019folding}, adapted from an anomalous on-site-symmetry SET introduced in Ref.~\cite{fidkowski2017realizing}.
We explicitly compute the $H^2$ obstruction class and show that this SET is indeed anomalous.

\section{Symmetric gapped boundary of SET order}
\label{sec:bdry-set}
In this section, we follow the formulation of SET orders~\cite{barkeshli2019symmetry} to introduce on-site global symmetry $G$ into an intrinsic topological order with a gapped boundary.
Additionally, we review the description of the resulting symmetric gapped boundary~\cite{cheng2020relative}.
In fact, the symmetry group $G$~\footnote{We assume the symmetry group $G$ is finite.} of a bulk SET state is not always compatible with its gapped boundary, as observed in various studies~\cite{bischoff2019spontaneous, meir2012module}.
Both studies propose three levels of obstructions by investigating $G$-equivariant module categories over a $G$-crossed fusion category to characterize the incompatibility.
From a more physical perspective, we elucidate these obstructions.
The first-level obstruction concerns whether a Lagrangian algebra characterizing the gapped boundary is invariant under the symmetry action.
We shall always assume the first-level obstruction vanishes.
We then define an obstruction function, called $H^2$ obstruction, for the second-level obstruction.
When the $H^2$ obstruction vanishes, we obtain the classification group of inequivalent symmetry actions on boundary anyon-condensation spaces, which we denote by boundary $U$-symbols.
Then, the boundary $U$-symbols can determine unconfined defects that can cross the boundary to become $G$-defects of a $G$-SPT order on the other side.
The $F$-symbols of the $G$-defects correspond to the third-level obstruction.
We leave further discussion of the third-level obstruction to when we study the particular case of mirror symmetry in Sec.~\ref{fif}.

\begin{figure}[H]
	\centering
	\includegraphics[width=0.45\textwidth,height=2.5cm]{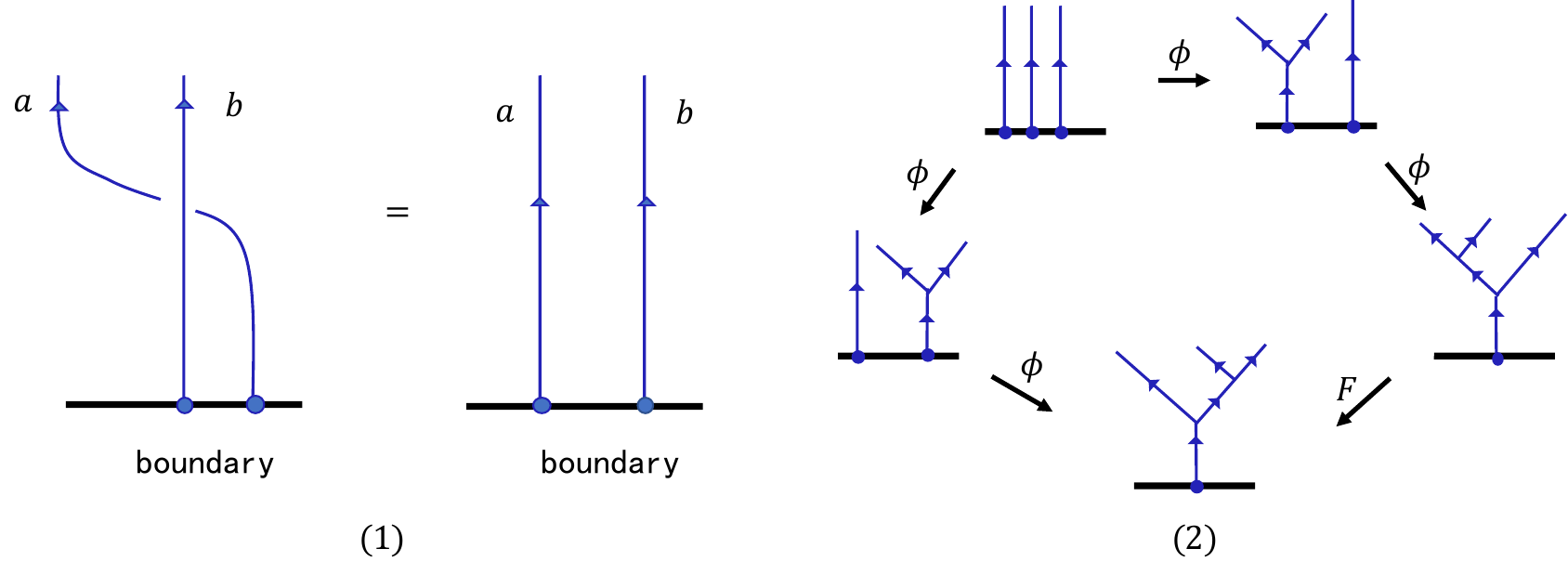}
	\caption{Coherent conditions of VLCs.}
	\label{fig1}
\end{figure}

For an intrinsic topological order $\cC$, we can investigate associated gapped boundaries through the mechanism of anyon condensation~\cite{eliens2014diagrammatics, bais2009condensate}.
A gapped boundary is described by a Lagrangian algebra in $\cC$.
This algebra is represented by a pair $(A, \phi)$, where $A$ is an object of $\cC$, and the VLCs $\phi$ originate from the algebra's multiplication~\cite{eliens2014diagrammatics, cong2016topological}.
Without ambiguity, we will use $A$ to denote both the object and the algebra itself. 
The object $A$ consists of all condensed anyons shown as $A=\bigoplus_a W_a a$, where the associated positive integer $W_a$ denotes the multiplicity of available condensation channels of condensed anyon $a$ on the $A$-type boundary. 
We note that the condensed anyons must be bosons with trivial topological spin,  and there must exist at least one condensed anyon $c$ such that the fusion channel satisfies $N^c_{ab} \geq 1$ for any two condensed anyons $a$ and $b$~\cite{bais2009condensate}.
In this paper, to simplify our analysis, we assume that the condensation channel of a condensed anyon is single (\emph{i.e.}, $W_a\leq1$).
Furthermore, we provide a physical perspective to elucidate the VLCs.
Considering a condensed anyon $a$, the corresponding dual anyon $\bar{a}$~\footnote{In a topological order, every anyon $a$ has unique dual anyon $\bar{a}$ which satisfies $N^{\bm{1}}_{a \bar{a}} = 1$ where $\bm{1}$ is the trivial anyon.} can also condense on the boundary~\cite{eliens2014diagrammatics}.
Since $N^{\bm{1}}_{a \bar{a}} = 1$, there are topological states that only involve the anyon pair $a$ and $\bar{a}$ within the bulk.
If the dual anyon $\bar{a}$ moves to the boundary and condenses to vacuum, additional topological states are created, with the anyon $a$ isolated within the bulk.
Therefore, we establish a series of condensation spaces that include the additional topological states, denoted as $V^a$, each associated with a basis $\ket{a,A}$ for each condensed anyon $a$.
For the whole system, we enlarge the topological space of the bulk by taking tensor products of condensation spaces $V^a$ with themselves and splitting spaces $V^{ab}_c$.
Based on a physical fact that condensing two bosons is equivalent to initially fusing them and subsequently condensing the resultant entity, the vectors of $V^a \otimes V^b$ can be expanded on $ \bigoplus_c  V^{ab}_c \otimes V^c $. 
The VLCs manifest as the expansion coefficients of bases,
\begin{gather}
	\ket{a,A}\otimes \ket{b,A}=\sum_{c,\alpha}\phi^{a b}_{c,\alpha} \ket{a, b; c,\alpha}\otimes \ket{c,A}.
\end{gather}   
There are two coherent conditions of VLCs diagrammatically depicted as Fig.~\ref{fig1}.
\noindent
That is,
\begin{gather}
	\phi^{a b}_{c,\alpha}=\sum_{\beta}\phi^{b a}_{c,\beta} [R^{ab}_c]_{\beta \alpha} \label{2}, \\
	\sum_{e,\alpha,\beta} \phi^{ab}_{e,\alpha; }\phi^{e c}_{d,\beta}[F^{abc}_d]_{(e,\alpha,\beta)(f,\mu,\nu)}=\phi^{b c}_{f,\mu}\phi^{a f}_{d,\nu} \label{3}.
\end{gather}
We obtain available VLCs by \Eqs{2}{3} which characterize a specific gapped boundary~\cite{eliens2014diagrammatics, cong2016topological}.

Now, we introduce an on-site symmetry $G$ into the whole system. 
The action of group elements on anyons is determined by homomorphism,
\begin{gather}
	[\rho] : G \longrightarrow \text{Aut}(\cC),
\end{gather} 
where $[\rho_{\bg}]$ is an equivalent class of maps for each $\bg \in G$. 
We select a representative $\rho_{\bg}$ in class $[\rho_{\bg}]$ and denote $\rho_{\bg}(a)$ as $^{\bg}a$.
Furthermore, we define associated symmetry action on splitting spaces,
\begin{gather}
    \rho_{\bg}(\ket{a, b; c, \mu}) = \sum_{\nu}[U_{\bg}({\ubg a}, {\ubg b}; {\ubg c})]_{\mu \nu}\ket{{\ubg a}, {\ubg b}; {\ubg c}, \nu}.	 	
\end{gather}
This action induces transformations of $F$-symbols and $R$-symbols, which characterize the topological order $\cC$. 
Besides, these symbols are required to be invariant under the symmetry action $\rho_{\bg}$. 
That is,
\begin{widetext}
\begin{align}
    \rho_{\bg}([F^{a b c}_d]_{(e, \alpha, \beta)(f, \mu, \nu)})={}& \sum_{\alpha \prime, \beta \prime, \mu \prime, \nu \prime} [U_{\bg}({\ubg a}, {\ubg b}; {\ubg e})]_{\alpha \alpha \prime}[U_{\bg}({\ubg e}, {\ubg c}; {\ubg d})]_{\beta \beta \prime}  [F^{{\ubg a}{\ubg b}{\ubg c}}_{\ubg d}]_{({\ubg e}, \alpha \prime, \beta \prime)({\ubg f}, \mu \prime, \nu \prime)} \notag \\
      {}&\times [U_{\bg}({\ubg b}, {\ubg c}; {\ubg f})^{-1}]_{\mu \mu \prime} [U_{\bg}({\ubg a}, {\ubg f}; {\ubg d})]_{\nu \nu \prime} = [F^{a b c}_d]_{(e, \alpha, \beta)(f, \mu, \nu)} \label{6}, \\
      \rho_{\bg}([R^{ab}_{c}]_{\mu \nu}) ={}& \sum_{\mu \prime \nu \prime}[U_{\bg}({\ubg b}, {\ubg a}; {\ubg c})]_{\mu \mu \prime}[R^{{\ubg a}{\ubg b}}_{\ubg c}]_{\mu \prime \nu \prime}[U_{\bg}({\ubg a}, {\ubg b}; {\ubg c})^{-1}]_{\nu \prime \nu}  = [R^{ab}_{c}]_{\mu \nu} \label{7}.
\end{align}
We assume that the bulk is non-anomalous. 
Then, there are phase factors $\eta_{a}(\bg, \bh)$ which characterize symmetry fractionalization and satisfy two coherent conditions,
\begin{gather}
	\frac{\eta_{a}(\bg,\bh)\eta_{b}(\bg,\bh)}{\eta_{c}(\bg,\bh)}\delta_{\alpha \beta}=\sum_{\mu,\nu}[U_{\bg \bh}(a,b;c)]_{\alpha \mu}[U_{\bg}(a, b;c)^{-1}]_{\mu \nu}[U_{\bh}({\ubbg a},{\ubbg b};{\ubbg c})^{-1}]_{\nu \beta}\label{8}, \\
	 \eta_{\ubbg a}(\bh, \bk)\eta_a(\bg,\bh \bk) = \eta_a(\bg \bh, \bk)\eta_a(\bg, \bh) \label{9}.
\end{gather}
\end{widetext}
The symmetry data $U, \eta$ symbols are obtained by \Eqss{6}{9} which characterize the bulk of the SET order except for symmetry defects.

Moving on to the boundary, the enlarged topological space involves condensation spaces.
As mentioned before, we assume that $A$ is invariant under symmetry action, which corresponds to the vanishing of the first level obstruction.
Consequently, we can define a symmetry action on these condensation spaces as
\begin{gather}
    \rho_{\bg} \ket{a,A}=U_{\bg}({\ubg a})\ket{{\ubg a},A},	
\end{gather}
where $\ubg A = A$.
Besides, this symmetry action on condensation spaces induces transformation of the VLCs and the VLCs should be invariant under the transformation.
That is, 
\begin{align}
    \rho_{\bg}({\phi}^{a b}_{c,\alpha})={}&\sum_{\beta} {\phi}^{{\ubbg a} {\ubbg b}}_{{\ubbg c}, \beta}[U_{\bg}(a, b;c)]_{\beta \alpha}\frac{U_{\bg}(c)}{U_{\bg}(a)U_{\bg}(b)}\notag \\ ={}& {\phi}^{a b}_{c,\alpha}	\label{11}.
\end{align}
The boundary $U$-symbols are derived from \Eq{11}. 
However, it is not the case that all solutions are physically meaningful. 
Considering the associated physical system, we define a physical state as $\ket{\Psi_{a_s,A}}$, which corresponds to the topological state $\ket{a,A}$. 
The subscript $s$ denotes the region $\cR_{s} $ where anyon $a$ is localized. 
Then, the symmetry action on physical states can be written as 
\begin{gather}
	R_{\bg}\ket{\Psi_{a_{s},A}}=U^{(s)}_{\bg}\rho_{\bg}\ket{\Psi_{a_{s},A}}=U^{(s)}_{\bg}U_{\bg}({\ubg a})\ket{\Psi_{{\ubg a}_{s},A}} \label{12},
\end{gather}
where $U^{(s)}_{\bg}$ is a local unitary operator which roughly acts within $\cR_{s} $. 
Furthermore, local operators $U^{(s)}_{\bg}$ satisfy projective multiplication relations
\begin{gather}
R_{\bg}U^{(s)}_{\bh}R_{\bg}^{-1}U^{(s)}_{\bg}\ket{\Psi_{a_s,A}}=\eta_{a}(\bg, \bh)U^{(s)}_{\bg \bh}\ket{\Psi_{a_s,A}}	\label{13}.
\end{gather}
It follows that
\begin{gather*}
 	R_{\bg}R_{\bh}\ket{\Psi_{{^{\overline{\bg \bh}}a}_{s}},A}=U_{\bg}(a)U_{\bh}(\ubbg a)\eta_{a}(\bg, \bh)U^{(s)}_{\bg \bh}\ket{\Psi_{a_s,A}}.
\end{gather*}
According to the group homomorphism $R_{\bg}R_{\bh}=R_{\bg \bh}$ , we derive
 \begin{gather}
 	\frac{U_{\bg \bh}(a)}{U_{\bg}(a)U_{\bh}(\ubbg a)}=\eta_a(\bg, \bh) \label{14},
 \end{gather} 
which is an additional constraint on boundary $U$-symbols $U_{\bg}(a)$. 
In reality, there may not exist suitable $U_{\bg}(a)$ satisfying the constraint, indicating that the introduced global symmetry $G$ is incompatible with the $A$-type boundary. 
In order to diagnose potential obstruction conveniently, we define a function 
\begin{align}
	O_a(\bg, \bh)=\eta^{-1}_a(\bg, \bh)\frac{U_{\bg \bh}(a)}{U_{\bg}(a)U_{\bh}(\ubbg a)} \label{15},
\end{align}
which corresponds to the second level obstruction.    
Here, we note that the quantity $O_a(\bg, \bh)$ is independent of the gauge choices of bulk. 
Specifically, there are two types of gauge transformations in bulk~\cite{barkeshli2019symmetry},
\begin{gather}
	\widetilde{\ket{a, b; c, \mu}} = \sum_{\nu} [\Gamma^{a b}_c]_{\mu \nu} \ket{a, b; c, \nu} \label{g1}, \\ 
	\breve{\rho}_{\bg}\ket{a, b; c, \mu} =\frac{\gamma_a(\bg)\gamma_b(\bg)}{\gamma_c(\bg)} \sum_{\nu}[U_{\bg}(a, b; c)]_{\mu \nu} \ket{a, b; c, \nu}\label{g2}.
\end{gather}
Apparently, both $\eta$-symbols and boundary $U$-symbols are invariant under the first type of gauge transformation(\ref{g1}). 
Under the second type of gauge transformation(\ref{g2}), bulk $U$-symbols and $\eta$-symbols are modified as
\begin{align}
    [\breve{U}_{\bg}(a, b; c)]_{\mu \nu} &= \frac{\gamma_a(\bg)\gamma_b(\bg)}{\gamma_c(\bg)}[U_{\bg}(a, b; c)]_{\mu \nu}, \\
	\breve{\eta}_a(\bg, \bh)&=\frac{\gamma_a(\bg \bh)}{\gamma_{\ubbg a}(\bh)\gamma_a(\bg)}\eta_a(\bg, \bh). 
\end{align} 
The modification of  $\breve{U}_{\bg}(a, b; c)$ leads to a corresponding alteration in the solution for $\breve{U}_{\bg}(a)$. 
Consequently, the quantity $O_a(\bg, \bh)$ is still invariant. 
According to \Eqs{8}{11}, we can derive
\begin{gather}
	O_a(\bg, \bh)O_b(\bg, \bh)=O_c(\bg, \bh) \label{16}.
\end{gather}
Moreover, \Eq{9} implies
\begin{gather}
\frac{O_{^{\bar{\bg}}a}(\bh, \bk)O_a(\bg,\bh \bk)}{O_a(\bg \bh, \bk)O_a(\bg, \bh)}=1 \label{17}
\end{gather}
so we can write $O_a(\bg, \bh)=M^*_{a \cO(\bg, \bh)}$ for {$\cO(\bg, \bh) \in \cA_{\cC}$ where $\cA_{\cC}$ is a set of abelian anyons in  $\cC$. 
Because of the trivial mutual statistical braiding between condensed anyons, we argue that $\cO(\bg, \bh) \in \cA_{\cC}/\cA_{A}$ where $\cA_{A}$ represents the set of condensed abelian anyons on an $A$-type boundary.
Then, \Eq{17} tells us $\cO \in Z^2_{\rho}[G,\cA_{\cC}/\cA_{A}]$.
In addition, we can modify the solutions $U_{\bg}(a)$ of \Eq{9} as 
\begin{gather*}
	\breve{U}_{\bg}(a) = U_{\bg}(a)M^*_{a\tau(\bg)},
\end{gather*}
where $\tau(\bg) \in C^1[G, \cA_{\cC}/\cA_{A}]$. 
Meanwhile, it changes the obstruction as 
\begin{gather}
	\breve{O}_{a}(\bg, \bh) = O_{a}(\bg, \bh)\frac{M^*_{a\tau(\bg \bh)}}{M^*_{a\tau(\bg)}M^*_{a{\ubg \tau}(\bh)}} \label{18}
\end{gather}
so we consider the modified $\breve{O}$ to be equivalent. 
Consequently, the inequivalent $O$ is characterized by 
\begin{gather*}
	\mathcal{O} \in H^2_{\rho}[G,\mathcal{A}_{\mathcal C}/\mathcal{A}_{A}].
\end{gather*}
Thus we refer to the quantity $O$ as $H^2$ obstruction.
If there are two groups of solutions labeled by $U_{\bg}(a), U^{\prime}_{\bg}(a)$, they may lead to different obstructions $O_{a}(\bg, \bh), O^{\prime}_{a}(\bg, \bh)$. 
We define $V_{\bg}(a) = \frac{U^{\prime}_{\bg}(a)}{U_{\bg}(a)}$. 
Applying \Eq{11}, we have $V_{\bg}(a)V_{\bg}(b)=V_{\bg}(c)$. 
It means $V_{\bg}(a) = M^*_{a\tau(\bg)}$ where $\tau(\bg) \in C^1[G, \mathcal{A}_{\cC}/\mathcal{A}_{A}]$. 
Then, the two obstructions are equivalent. 
As a result, the inequivalent $H^2$ obstruction is determined by symmetry action $\rho$.

Finally, we discuss the classification of symmetry action on condensation spaces given by $U_\bg(a)$, when the obstruction $O$ vanishes.
Without loss of generality, we can select $O_a(\bg, \bh)=1$ in the trivial equivalence class $\cO = [0]$. 
Then we can obtain meaningful $U_{\bg}(a)$ by \Eqs{11}{14}. 
Furthermore, $U_{\bg}(a)$ can be twisted by factors $\chi_{\bg}(a)$ which satisfies
\begin{equation}
\begin{gathered}
	\chi_{\bg}(a)\chi_{\bg}(b)=\chi_{\bg}(c), \\
	\chi_{\bg \bh}(a)=\chi_{\bg}(a)\chi_{\bh}(^{\bar{\bg}}a).
\end{gathered}
\end{equation}
Again we can write $\chi_{\bg}(a)=M^*_{a\mathfrak{v}(\bg)}$ where $\mathfrak{v} \in Z^1_{\rho}[G, \mathcal{A}/\mathcal{A}_{A}]$. 
However, there is redundancy for $U_{g}(a)$ based on the gauge transformation for condensation spaces $V^a$,
\begin{gather*}
	\widetilde{\ket{a,A}} = \gamma_{a}\ket{a,A}.
\end{gather*} 
If we have chosen a proper gauge to fix $\phi^{ab}_{c,\alpha}$, then $\gamma_{a}\gamma_{b}=\gamma_{c}$. 
Therefore, we argue that the solution $U_{\bg}(a)$ is equivalent to $\breve{U}_{\bg}(a)=U_{\bg}(a)\frac{\gamma_{a}}{\gamma_{^{\bar{\bg}}a}}$. 
The inequivalent solution $U_{g}(a)$ is classified by 
\begin{gather*}
    H^1_{\rho}[G,\mathcal{A}_{\mathcal{C}}/\mathcal{A}_{A}],
\end{gather*}
which partially characterizes a symmetric gapped boundary.

\section{solution to anomalous case} \label{thi} 
In this section, we focus on 2D SET orders with an anomalous boundary, indicated by a nonvanishing $H^2$ obstruction.
Generally, an anomalous topological phase cannot be realized alone, and must be realized on the surface of a higher-dimensional topological phase that can cancel out the anomaly~\cite{chen2015anomalous,fidkowski2017realizing,barkeshli2018time}.
The $H^2$ obstruction appears at the 1d boundary, so we intend to cancel the anomaly by constructing a proper child 2D SET order on the other side of the boundary.
Motivated by a specific example detailed in Ref.~\cite{chen2015anomalous}, we provide a general method for constructing such child phases.
Moreover, we prove that the obtained child phase can be integrated with the original SET order along the anomalous boundary to make the whole system anomaly-free.
As a result, the anomalous boundary turns into an interface between two SET orders.

We start from an intrinsic topological order $\cC$ enriched by global symmetry $G$ with a gapped boundary characterized by algebra $A$.
Here, we suppose that the algebra $A$ is invariant under symmetry action.
However, the gapped boundary is found to be incompatible with the group $G$ which   implies the presence of a non-trivial $H^2$ obstruction.
Then, we construct a condensible subalgebra $B = \oplus_{b}b$ of $\cC$ by selecting $A$-type condensed anyons which satisfy $O_{b}(\bg, \bh)=1$.
We note that \Eq{16} guarantees that the condensable subalgebra $B$ makes sense.
Apparently, this subalgebra $B$ is invariant under the symmetry action as well. 
Furthermore, we can obtain a desired child topological order $\cD$ by condensing $B$ on the interface. 
We denote anyons in $\cD$ by $i,j,k, \dots$
\begin{figure}[H]
	\centering
	\includegraphics[width=0.45\textwidth,height=3cm]{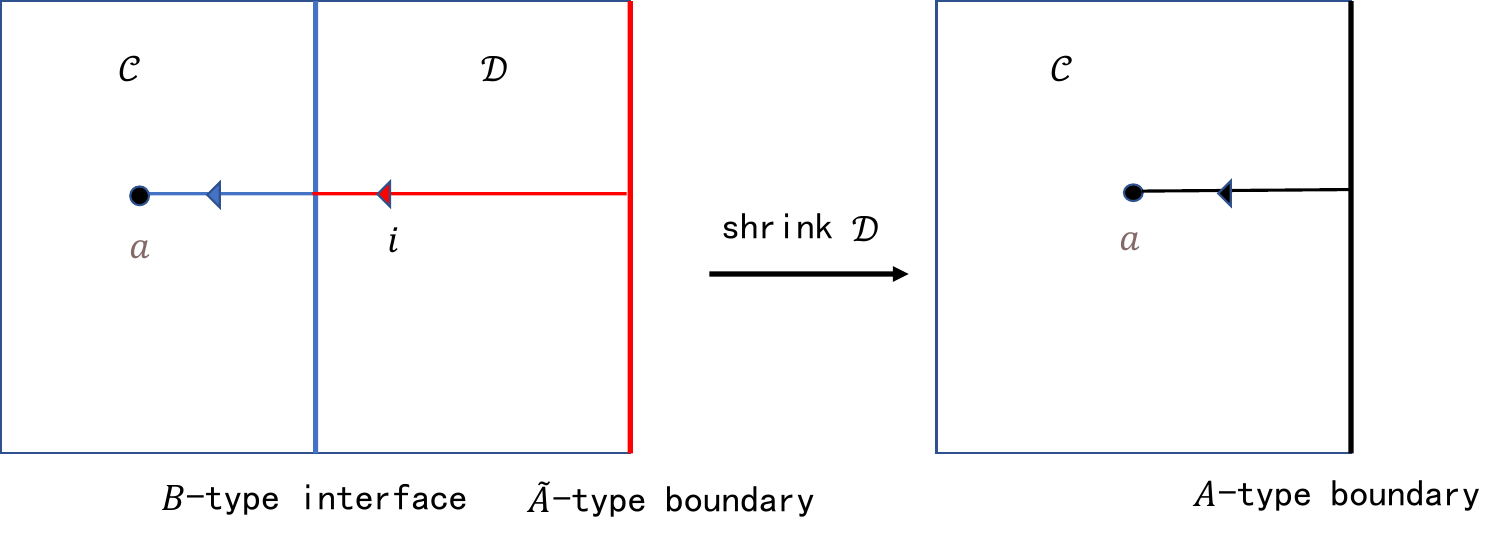}
	\caption{Two-step anyon condensation}
	\label{fig2}
\end{figure}
\noindent
Using the concept of a two-step anyon condensation~\cite{KONG2014436,cong2017defects}, we consider a special gapped boundary of $\cD$ characterized by Lagrangian algebra $\tilde{A}$ which is entirely determined by the two algebras $A$ and $B$,
\begin{gather}
    \tilde{A}= \oplus_{i} i.
\end{gather}
Here, condensed anyon $i$ of an $\tilde{A}$-type boundary can be determined by the lifting map $l$ of $B$-type interface,
\begin{gather}
    l(i) = \oplus_{a} W_{ai} a, 	
\end{gather}
where $a$ is condensed anyon of $A$-type boundary and $W_{ai}\leq1$.
Especially, the lifting of trivial anyon $1_{\cD}$ of $\cD$ is 
\begin{gather}
    l(1_{\mathcal{D}}) = B. 	
\end{gather}
In this case, the two-step anyon condensation is illustrated in Fig.~\ref{fig2}.
The $B$-type interface and $\tilde{A}$-type boundary can fuse into the $A$-type boundary if we shrink child topological order $\cD$. 
Besides, on the $\tilde{A}$-type boundary, there is a unique condensed anyon $i$ which can be lifted to a given condensed anyon $a$ on the $A$-type boundary.
Given a condensed anyon $i$ on the $\tilde{A}$-type boundary, if two condensed anyons $a_{1}$ and $a_{2}$ of $A$-type boundary are both lifts of $i$, then there must be a condensed anyon $b$ such that $N^{a_{2}}_{a_{1}b}>0$ on the $B$-type interface. 
It implies that
\begin{align}
	O_{a_2}(\bg, \bh) = O_{a_1}(\bg, \bh).
\end{align}
Thus, we can denote the independent obstruction $O_{a}(\bg, \bh)$ by $O_{i}(\bg, \bh)$ without ambiguity. 
According to the commutativity between lifting map and fusion in anyon condensation theory, we have $O_i(\bg, \bh)O_j(\bg, \bh)=O_k(\bg, \bh)$. 
Consequently, the $H^2$ obstruction is in a subgroup of $H^2_{\rho}[G,\cA_{\cC}/\cA_{A}]$. 
Specifically, it is 
\begin{gather}
      H^2_{\rho}[G, \mathcal{A}_{\mathcal{D}}/ \mathcal{A}_{\tilde{A}} ]	.
\end{gather}
Furthermore, we introduce the on-site symmetry $G$ into the child topological order $\cD$ with induced symmetry action on anyons in $\cD$. 
There may be nontrivial $H^2$ obstruction $\mathfrak{o}\in H^2_{\rho}[G, \cA_{\cD}/\cA_{\tilde{A}} ]$ on the $\tilde{A}$-type boundary which can be used to match the $H^2$ obstruction on the $A$-type boundary.
Following the formulation discussed in Sec.~\ref{sec:bdry-set}, the obstruction $\mathfrak{o}$ is defined as 
\begin{gather}
M^*_{i\mathfrak{o}(\bg, \bh)}	=\tilde{\eta}_{i}(\bg, \bh)^{-1}\frac{U_{\bg \bh}(i)}{U_{\bg}(i)U_{\bh}(^{\bar{\bg}}i)},
\end{gather}
where the phase factor $\tilde{\eta}_i(\bg, \bh)$ characterizes symmetry fractionalization in the bulk of the child SET order. 
It implies that we can adjust $\mathfrak{o}$ by $\tilde{\eta}_i(\bg, \bh)$ to match $\cO$. 
Consequently, we obtain a child SET order with proper $\eta$-symbols where we ignore the data of symmetry defects.

Now, we investigate the interface between the two SET orders to show that the obtained child SET cancels the $H^2$ obstruction.
If we fold the child SET order, then the interface will turn into a symmetric gapped boundary of the induced bilayer system.
As a result, we apply the formulation of a symmetric gapped boundary discussed in Sec.~\ref{sec:bdry-set} to study the $B$-type interface.
In this case, the condensed spaces are denoted as $V^a_i$ with a corresponding basis $\ket{a; i}$.
The basis $\ket{a; i}$ corresponds to the fact that condensed anyon $a$ on an $A$-type boundary can cross the $B$-type interface to transform into condensed anyon $i$ on an $\tilde{A}$-type boundary.
The associated VLCs are defined as
\begin{align}
   \ket{a;i} \otimes \ket{b; j} ={}& \notag \\
     \sum_{c, k,\alpha,\beta} \phi^{a b; k,\beta}_{c,\alpha; i j} {}& \ket{a, b; c,\alpha}	\otimes \ket{c; k}\otimes \ket{k,\beta;i,j}.
\end{align}
Similarly, there are two coherent equations for the VLCs.
Besides, we define the induced symmetry action on the condensed space $V^a_i$ as
\begin{gather}
       \rho_{\bg}\ket{a; i} = U_{\bg}(a; i)\ket{\ubg a; {\ubg i}}.	
\end{gather}
As a result, the VLCs of the $B$-type interface should be invariant under the induced symmetry transformation,
\begin{widetext}
\begin{gather}
 	\phi^{a b; k,\gamma}_{c,\alpha; i j}=\sum_{\beta,\sigma}\phi^{b a; k,\sigma}_{c,\beta; j i}[R^{ab}_c]_{\beta \alpha}[R^{ij}_k]^{-1}_{\sigma \gamma},\\
    \sum_{e,\alpha,\beta, m,\sigma,\rho} \phi^{ab; m,\sigma}_{e,\alpha; ij} \phi^{e c; l,\rho}_{d,\beta; m k}[F^{abc}_d]_{(e,\alpha,\beta)(f,\mu,\nu)}[F^{ijk}_l]^{-1}_{(m,\sigma,\rho)(n,\lambda,\tau)}= \phi^{b c; n,\lambda}_{f,\mu; j k} \phi^{a f; l,\tau}_{d,\nu; i n},\\
    \rho_{\bg}(\phi^{a b; k,\sigma}_{c,\alpha; i j})=\sum_{\beta,\rho}\phi^{^{\bar{\bg}}a ^{\bar{\bg}}b; ^{\bar{\bg}}k,\rho}_{^{\bar{\bg}}c,\beta; ^{\bar{\bg}}i ^{\bar{\bg}}j}[U_{\bg}(a, b; c)]_{\beta \alpha}[U_{\bg}(i, j; k)]^{-1}_{\rho \sigma}U_{\bg}(c; k)U^{-1}_{\bg}(a; i)U^{-1}_{\bg}(b; j)= \phi^{a b; k,\sigma}_{c,\alpha; i j} \label{29}.
\end{gather}
\end{widetext}
According to the \Eqs{2}{3}, we derive $\phi^{a b; k,\sigma}_{c,\alpha; i j}= {\phi^{a b}_{c,\alpha}}{[\phi^{i j}_{k,\sigma}}]^{-1}$. 
Plugging \Eq{11} into \Eq{29}, we derive 
\begin{gather}
    U_{\bg}(a; i)= \frac{U_{\bg}(a)}{U_{\bg}(i)}	.
\end{gather} 
For physical system, we consider the action of symmetry operator $R_{\bg}$ on a physical state $\ket{\Psi_{a_{s_1}; i_{s_2}}}$, which corresponds to topological state $\ket{a; i}$.
Similarly, the equation $R_{\bg}R_{\bh}\ket{\Psi_{a_{s_1}; i_{s_2}}}=R_{\bg \bh}\ket{\Psi_{a_{s_1}; i_{s_2}}}$ forces 
\begin{gather}
    \frac{U_{\bg \bh}(a; i)}{U_{\bg}(a; i)U_{\bh}({^{\bar{\bg}}a}; {^{\bar{\bg}}i})}=\frac{\eta_a(\bg, \bh)}{\tilde{\eta}_i(\bg, \bh)}.	
\end{gather}
This equation determines the proper $\eta$-symbols in child SET order.
Plugging the solution of $U_{\bg}(a; i)$ into it, we derive 
\begin{gather}
    \mathcal{O} = \mathfrak{o},	
\end{gather}
which is exactly the result of obstruction vanishing.

\section{Mirror symmetry enriched topological phase} \label{for}
In this section, we study mirror symmetry enriched topological phases.
We apply the folding approach~\cite{qi2019folding} to avoid challenging the nonlocality of mirror symmetry. 
Different from gauging $\mathbb{Z}_2$ interlayer symmetry in a bilayer system and studying various symmetric interfaces by anyon condensation~\cite{qi2019folding}, we derive symmetry data by the formulation discussed in Sec.~\ref{sec:bdry-set}.
Besides, we find that there is potential $H^2$ obstruction on the mirror axis. 
When the $H^2$ obstruction vanishes, we study the t'Hooft anomaly of mirror SET orders corresponds to the third level obstruction on the gapped boundary of bilayer system.

Starting from an intrinsic topological order $\cC$, we fold it along the mirror axis to produce a bilayer system $\cC \boxtimes \cC^{\text{rev}} $ where $\cC^{\text{rev}}$ is reverse of $\cC$. 
The mirror axis turns into a gapped boundary. 
Due to folding, the normal direction of a layer is inverted.
Thus, there is one-to-one correspondence between anyon $a^{\text{rev}} \in \cC^{\text{rev}}$ and anyon $a \in \cC$.
Here, $a^{\text{rev}}$ is considered to be the reverse counterpart of $a$, with a topological spin relation $\theta_{a}= \theta^{*}_{a^{\text{rev}}}$. 

In the bilayer system, the mirror symmetry $\mathbb{Z}^{\mb}_2=\{\be, \mb \}$ is converted into on-site symmetry $\mathbb{Z}^{\bM}_2 = \{\be, \bM \}$. 
It is known that a non-anomalous SET order is classified by three levels data $\rho, \mathfrak{w}, \alpha$,
\begin{equation}
\begin{gathered}
	\rho: \mathbb{Z}^{\bM}_2 \rightarrow \text{Aut}(\mathcal{C} \boxtimes \mathcal{C}^{\text{rev}}),\\
	\mathfrak{w} \in H^2_{\rho}[\mathbb{Z}^{\bM}_2, \mathcal{A}_{\mathcal{C} \boxtimes \mathcal{C}^{\text{rev}}}],\\
	\alpha \in H^3[\mathbb{Z}^{\bM}_2, U(1)].
\end{gathered}
\end{equation}  
In this case, the map of $\mathbb{Z}^{\bM}_2$ is induced from a map of $\mathbb{Z}^{\mb}_2$ shown as
\begin{gather}
	\rho : \mathbb{Z}^{\mb}_2 \rightarrow \text{Aut}^*(\mathcal{C}).
\end{gather} 
Particularly, the nontrivial permutation $\rho_{\mb}$ is an anti-auto-equivalence of $\cC$ included in $\text{Aut}^*(\cC)$.    
It implies $\theta_{\ubm a} = \theta^*_a$.
Therefore, we can relabel anyon $\ubm a^{\text{rev}} \in \cC^{\text{rev}}$ by $a$.
We note that $\cC$ must be equivalent to $\cC^{\text{rev}}$, which is a constraint on topological orders with mirror symmetry. 
Applying new notation, we denote an anyon of the bilayer system as $(a, b)$, so the nontrivial symmetry action of $\mathbb{Z}^{\bM}_2$ turns into layer exchange, 
\begin{gather}
	\rho_{\bM}: (a, b) \rightarrow (b, a).
\end{gather}
The specified $\rho$ implies non-anomalous bulk~\cite{gannon2019vanishing} and trivial group $H^2_{\rho}[\mathbb{Z}^{\bM}_2,\cA_{\cC \boxtimes \cC^{\text{rev}}}]$ which characterizes the classification of symmetry fractionalization in bulk.
Besides, we can produce different $\mathbb{Z}^{\bM}_2$ SET order by stacking a different $\mathbb{Z}^{\bM}_2$-SPT phase characterized by a torsor $\alpha \in H^3[\mathbb{Z}^{\bM}_2, U(1)]$.
More specifically, the torsor $\alpha$ can modify the topological spin of $\mathbb{Z}_2^{\bM}$ symmetry defects denoted as $x_{\bM}$ where $x \in \cC$,                   
\begin{gather}
    \tilde{\theta}^2_{x_{\bM}} = \alpha(\bM, \bM, \bM) {\theta}^2_{x_{\bM}}.
\end{gather}
However, only the $\mathbb{Z}^{\bM}_2$ SET order with topological spin of symmetry defect $\theta^2_{x_{\bM}} = \theta_{x}$ corresponds to the original mirror SET order~\cite{qi2019folding}.
Therefore, we focus on the selected $\mathbb{Z}^{\bM}_2$ SET order below.

As analyzed above, symmetry data are uniquely determined when we choose proper gauge in bulk. 
It is shown as
\begin{equation}
\begin{gathered}
   [U_{\bg}((a_1,a_2),(b_1,b_2);(c_1,c_2))]_{\mu \nu} = \delta_{\mu \nu},	\\
     \eta_{(a, b)}(\bg, \bh) = 1.
\end{gathered}
\end{equation}

Next, we move on to the gapped boundary, which encodes all information of the mirror SET~\cite{qi2019folding}.
In our notation, the associated Lagrangian algebra is expressed as
\begin{gather}
    A= \oplus_{a}a_A,
\end{gather}
where we label the condensed anyon $(a, \bar{\rho}_{\mb}(a))$ by $a_A$. 
Then the nontrivial \Eqs {11}{15} are 
\begin{gather}
    \phi^{a_A b_A}	_{c_A, \alpha} =  \phi^{{\bha}_A{\bhb}_A}_{{\bhc}_A; \alpha} \frac{U_{\bM}(c_A)}{U_{\bM}(a_A)U_{\bM}(b_A)} \label{41}, \\
    O^{-1}_{a_A}(\bM, \bM) = U_{\bM}(a_A)U_{\bM}({\bha}_A) \label{42}.
\end{gather}
There is potential $H^2$ obstruction which is characterized by an element in
\begin{gather}
	H^2_{\rho}[\mathbb{Z}^{\bM}_2, \cA_{\cC \boxtimes \cC}/\cA_A ] =H^2_{\rho}[\mathbb{Z}^{\mb}_2, \cA_{\cC}].
\end{gather}
Consequently, we have $O_{a_A}(\bM, \bM) = M^*_{a_A \mathcal{O}(\bM, \bM)}=M^*_{a\mathcal{O}(\mb, \mb)}$. 

Next, we consider the anomaly-free cases.
The inequivalent solutions $U_{\bM}(a_A)$ distinguish different symmetric gapped boundaries.
Furthermore, the classification group associated with the solution $U_{\bM}(a_A)$ is 
\begin{gather}
    H^1_{\rho}[\mathbb{Z}^{\bM}_2, \cA_{\cC \boxtimes \cC}/\cA_A ]	=H^1_{\rho}[\mathbb{Z}^{\mb}_2, \cA_{\cC}].
\end{gather}
Due to the particularity of mirror symmetry, we have 
\begin{align}
	H^1_{\rho}[\mathbb{Z}^{\mb}_2, \cA_{\cC}] = H^2_{\bar{\rho}}[\mathbb{Z}^{\mb}_2, \cA_{\cC}],
\end{align}
which is consistent with the results of Ref.~\cite{qi2019folding}.
The classification group of boundary $U$-symbols corresponds to the classification group of symmetry fractionalization in on-site SET orders exactly.
As a result, we claim that the boundary $U$-symbols characterize the symmetry fractionalization for mirror SET orders.
  
Next, we study the potential t'Hooft anomaly which can be canceled out by sticking to a nontrivial $\mathbb{Z}_2$-SPT along the boundary. 
On the symmetric boundary, we consider a defect $x_{\bM}$ that belongs to a $\bM$-local module of algebra $A$~\cite{bischoff2019spontaneous}.
As a result, it implies that the defect $x_{\bM}$ should satisfy the associated $\bM$-local condition illustrated in Fig.~\ref{f3}.
\begin{figure}[H]
	\centering
	\includegraphics[width=0.4\textwidth,height=3.5cm]{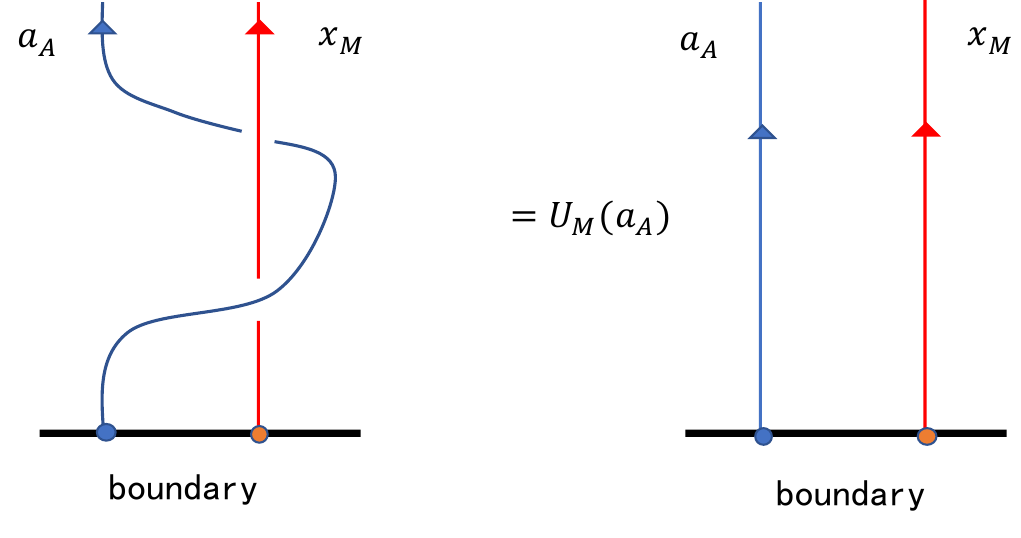}
	\caption{$\bM$-local condition.}
	\label{f3}
\end{figure}
\noindent
That is, 
\begin{gather}
    \sum_{\alpha, \beta} \mu^{{\bha}_A x_{\bM}}_{y_{\bM}; \alpha} [R^{x_{\bM} {\bha}_A}_{y_{\bM}}]_{\alpha \beta} [R^{a_A x_{\bM}}_{y_{\bM}}]_{\beta \gamma}
    	 = U_{\bM}(a_A)\mu^{a_A x_{\bM} }_{y_{\bM}; \gamma} \label{47}.	
\end{gather}
where $\mu$-symbols are defined by module actions of the $\bM$-local module~\cite{bischoff2019spontaneous}.
From a physical perspective, it implies that the defect $x_{\bM}$, referred to as unconfined, can cross the symmetric gapped boundary to become a $\mathbb{Z}^{\bM}_2$-defect of $\mathbb{Z}^{\bM}_2$ SPT order denoted as $M$.
Therefore, we have
\begin{align}
	\theta^2_{x_{\bM}}=\theta^2_M.
\end{align}
For symmetry defects, there is a Heptagon consistency equation~\cite{barkeshli2019symmetry}.
We write down the nontrivial one in the $\mathbb{Z}^{\bM}_2$ SPT order,
\begin{gather}
   [R^{MM}_{\bm{1}}]^{-1}F^{MMM}_M[R^{MM}_{\bm{1}}]^{-1}=F^{MMM}_M F^{MMM}_M,
 \end{gather}
where we choose a gauge to leave $\eta_{M}(\bg, \bh)=1$ and denote trivial anyon as ${\bm{1}}$.
Besides, we have a ribbon property of $R$-symbols,
\begin{align}
	R^{MM}_{\bm{1}}R^{MM}_{\bm{1}} = \frac{1}{\theta^2_{M} }.
\end{align} 
It follows that
\begin{gather}
    F^{MMM}_M = \theta_{x}. 
    \label{50}
\end{gather}
Because $F$-symbols satisfy the Pentagon equation, we have $F^{MMM}_M= \pm 1$, which characterizes inequivalent $\mathbb{Z}^{\bM}_2$ SPT orders.
There is a definition of the t'Hooft anomaly indicator of the mirror SET order proposed in Ref.~\cite{qi2019folding},
\begin{gather}
	\eta = \theta_{x}. 
\end{gather}
More specifically, $\eta=+1$ corresponds to non-anomalous case and $\eta=-1$ corresponds to anomalous case.
It exactly corresponds to the result in \Eq{50}. 
Besides, boundary $U$-symbols can be modified by a torsor $\mathfrak{v} \in H^1_{\rho}[\mathbb{Z}^{\mb}_2, \mathcal{A}_{\mathcal{C}}]$,
\begin{gather}
	\tilde{U}_{\bg}(a_A)=U_{\bg}(a_A) M^*_{a \mathfrak{v}(\bg)}.
\end{gather}
It may further alter t'Hooft anomaly indicator $\eta$. 
Considering \Eq{47}, we find that the deformed unconfined defect can be selected as $\tilde{x}_{\bM}$ where $N^{\tilde{x}}_{x\mathfrak{v}(\mb)}=1$. 
Due to the relation $S_{b_{\bM}(a, a)}=S_{ba}$~\cite{qi2019folding}, the relative anomaly indicator $\eta_r$ can be derived,
\begin{align}
	\eta_r=\theta_{\mathfrak{v}(\mb)}U_{\bM}(\mathfrak{v}(\mb)_A) \label{53}.
\end{align}

The above discussion of the t'Hooft anomaly is based on an assumption that the desired defect $x_{\bM}$ satisfying \Eq{47} exists.
However, bilayer $\mathbb{D}_{16}$ gauge theory is a counterexample which was studies in Ref.~\cite{qi2019folding}.
In our formulation, this case is regarded as an anomalous phase with $H^2$ obstruction, which will be discussed in Sec.~\ref{sec:d16}.
In fact, the $\bM$-local module of algebra $A$, which includes the desired defect $x_{\bM}$ must exist if the $H^2$ obstruction vanishes~\cite{bischoff2019spontaneous}.

\section{Realization of anomalous mirror SET orders with $H^2$ obstruction}\label{fif}                                    
In this section, we argue that mirror SET states with an $H^2$ obstruction can be realized as the surface topological order of a 3D bulk state.
Applying the approach developed in Sec.~\ref{thi}, we find a 2D child SET orders to cancel out the $H^2$ obstruction along the boundary.
Moreover, 2D child SET orders can be used to construct 3D mirror SET orders by layered construction~\cite{zhao2022string,jian2014layer,gaiotto2019condensations}.
Then, the 2D anomalous mirror SET orders can localize on surface of constructed 3D mirror SET orders.

In a bilayer system, considering an anomalous $A$-type boundary, we select condensed anyons to construct subalgebra $B = \oplus_{b_A}b_A$ by the condition $O_{b_A}(\bM, \bM)=1$, which is essential for deriving desired child topological order.
Particularly, the cocycle condition of $\cO(\bg, \bh)$ results,
\begin{gather}
    O_{\cO(\mb, \mb)_A}(\bM, \bM) = 1.
\end{gather}
Thus there is a nontrivial condensed anyon $\cO(\mb, \mb)_A$ in subalgebra $B$ at least.
Crossing the $B$-type interface, there are a series of anyons $(\cO(\mb, \mb), 1_{\cC}), (\cO^2(\mb, \mb), 1_{\cC}) \dots$ survived in child topological order $\cD$ where we denote trivial anyon of single-layer intrinsic topological order $\cC$ as $1_{\cC}$.
If an integer $n$ is the minimum integer that satisfies the condition $\cO^n(\mb, \mb) = 1_{\cC}$, the mutual braiding statistics between anyon $d$ of $\cD$ and $(\cO(\mb, \mb), 1_{\cC})$ must be expressed as $e^{\frac{2k\pi i}{n}}$ where $k = 1,2,\dots,n$. 
As a result, when $\theta_{\cO(\mb, \mb)}=1$, we can regard $(\cO(\mb, \mb), 1_{\cC})$ as $\mathbb{Z}_n$ charge in $\cD$.
When $\theta_{\cO(\mb, \mb)}=-1$, $n$ must be an even integer so we can regard $(\cO(\mb, \mb), 1_{\cC})$ as $\mathbb{Z}^f_n$ charge in $\cD$ where $\mathbb{Z}^f_n$ is $\mathbb{Z}^f_2$ central extension of $\mathbb{Z}_{n/2}$. 
Here, $\mathbb{Z}^f_2$ is fermion parity symmetry. 
Furthermore, we define a subset $\cE \subset \cC$ as
$\{1_{\cC}, \cO(\mb, \mb),\dots, \cO^{n-1}(\mb, \mb) \}$ with natural fusion structure inherited from $\cC$.
The centralizer of $\cE$ is
\begin{gather}
    \cE' = \{b|b\in \cC, M^*_{b \cO(\mb, \mb)}=1  \}.	
\end{gather}
In the case that $\cC$ is non-degenerate which is typically studied in physics, we have~\cite{etingof2015tensor}
\begin{gather}
    \text{Dim}(\mathcal{E})\text{Dim}(\mathcal{E}')=\text{Dim}(\mathcal{C}),
\end{gather}
so we directly obtain $\text{FPdim}(B) = \text{Dim}(\cC)/n$.
Moreover, we have~\cite{KONG2014436}
\begin{gather}
	\text{Dim}(\mathcal{D}) = \frac{\text{Dim}(\mathcal{C} \boxtimes \mathcal{C})}{\text{FPdim}(B)^2} = n^2.
\end{gather}
Besides, anyons $b_A$ that satisfy the condition $O_{b_A}(\bM, \bM) \neq 1 $ can survive as nontrivial anyon in $\cD$.
Since $\theta_{b_A}=1$, these survived anyons can be regarded as pure fluxes in $\cD$. 
Hence, we argue that the desired child topological order $\cD$ is $\mathbb{Z}_n$ gauge theory,
\begin{gather}
    \mathcal{D} = \text{Rep}(D(\mathbb{Z}_n))
    \label{56},	
\end{gather}
where $D(\mathbb{Z}_n)$ is non-twisted quantum double of $\mathbb{Z}_n$~\cite{kitaev2003fault}.
Next, we introduce $\mathbb{Z}^{\bM}_2$ into the child topological order $\cD$.
In this case, the map $\rho : \mathbb{Z}_2^{\bM} \rightarrow \text{Aut}(\cD)$ is induced by the parent SET order.
However, the symmetry fractionalization is classified by 
\begin{gather}
    H^2_{\rho}[\mathbb{Z}^{\bM}_2, \cA_{\cD}]	 = H^2_{\rho}[\mathbb{Z}^{\bM}_2, \cA^{\text{charge}}_{\cD}] \times H^2_{\rho}[\mathbb{Z}^{\bM}_2, \cA^{\text{flux}}_{\cD}]	.	
\end{gather}
Besides, it's known that $(\cO(\mb, \mb), 1_{\cC}) \in \cA^{\text{charge}}_{\cD}$.
As a result, we can always cancel out the $H^2$ obstruction by adjusting symmetry fractionalization of the child SET order.

\begin{figure}[H]
	\centering
	\includegraphics[width=0.4\textwidth,height=0.24\textheight]{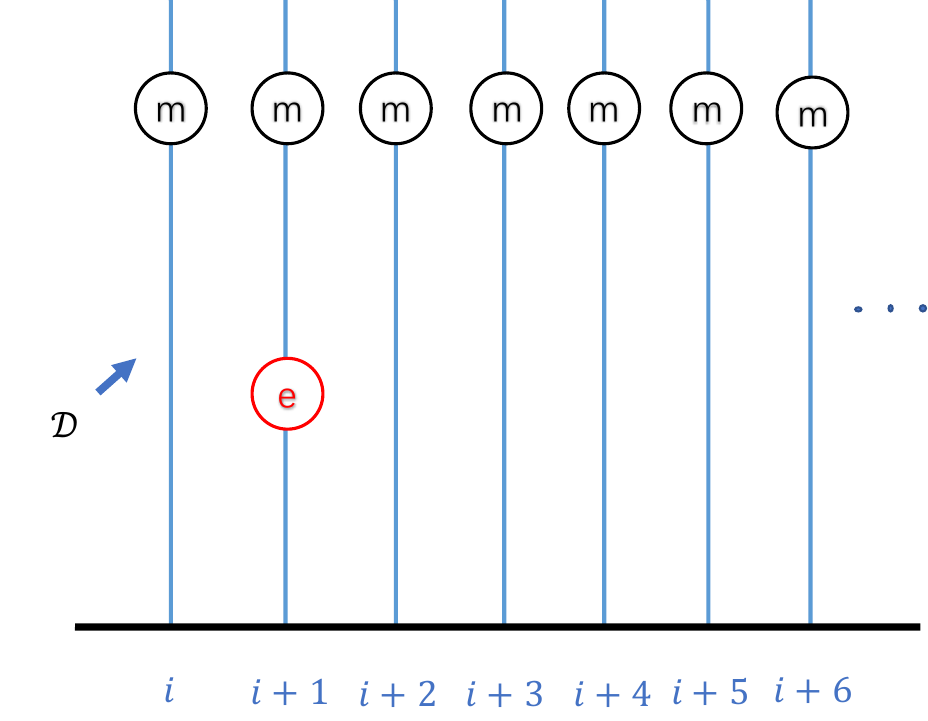}
	\caption{Layered construction of 3D $\mathbb{Z}_n$ gauge theory.}
	\label{4}
\end{figure}
Now we consider 3D $\mathbb{Z}_n$ gauge theory, which can be understood by layered construction~\cite{zhao2022string,jian2014layer,gaiotto2019condensations}.
As depicted in Fig.~\ref{4}, we stack consecutive layers of $\cD$ and condense paired charge excitations between adjacent layers.
These paired charge excitations are labeled by $e_i \otimes \bar{e}_{i+1}$ and are represented as red circles within the figure.
Thus charge excitations in 2D layers can jump into nearby layers as a result of anyon condensation.
It implies that charge excitations in 2D layers become unconfined $\mathbb{Z}_n$ gauge charges of the 3D bulk.
In this case, when $\theta_{\cO(\mb, \mb)}=1$, all $\mathbb{Z}_n$ gauge charges are bosons~\cite{lan2018classification}.
Conversely, for $\theta_{\cO(\mb, \mb)}=-1$, $\mathbb{Z}^f_n$ gauge charges encompass both fermions and bosons~\cite{lan2019classification}.
The configuration of flux excitations, depicted as black circles along a line, can be considered as flux string excitations within the 3D bulk.
These excitations correspond to $\mathbb{Z}_n$ gauge fluxes.
In the multiple layers system, we denote the $\cD$ localized at mirror plane as $\cD_m$.
Additionally, we pair two layers that are mirror symmetric to create a bilayer system, denoted as $\cD_{m-i} \boxtimes \cD_{m+i}$. 
Then we introduce mirror symmetry into the 3D $\mathbb{Z}_n$ gauge theory which is considered as comprising multiple layers.
Consequently, mirror symmetry transforms into layer-exchange symmetry for the bilayer subsystems and manifests as on-site symmetry at the mirror plane $\cD_m$.
We note that symmetry data for bilayer subsystems are trivial.
Furthermore, we adjust symmetry fractionalization of $\cD_m$ to cancel out $H^2$ obstruction that appears on the mirror axis of single layer $\cC$.
Consequently, we can localize the single layer 2D anomalous mirror SET at the boundary of the 3D mirror SET through a series of consecutive T-junctions, as illustrated in Fig.~\ref{5}.
These T-junctions are gapped interfaces, characterized by condensible subalgebra $B$ in folding geometry. 
Especially, the mirror plane is aligned with the mirror axis of $\cC$ to preserve the mirror symmetry. 
\begin{figure}[H]
	\centering
	\includegraphics[width=0.4\textwidth,height=4.5cm]{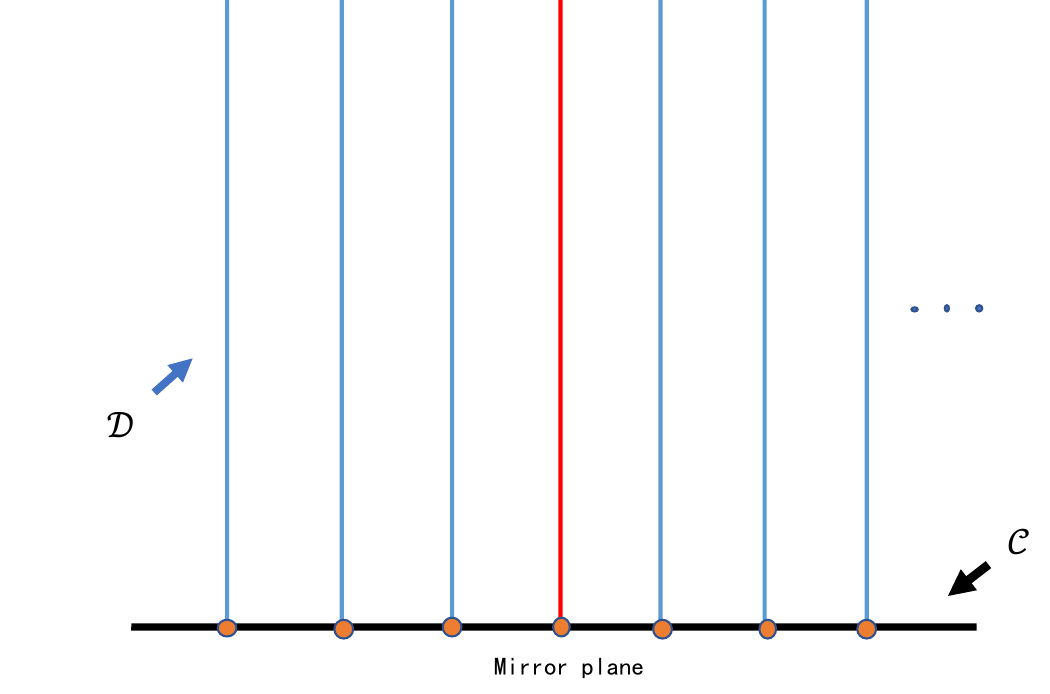}
	\caption{Anomalous 2D mirror SET order localized on the surface of 3D mirror SET order.}
	\label{5}
\end{figure}

\section{Correspondence between mirror SET and time reversal SET}
\label{sec:correspondence}
\begin{table*}[ht]
\centering
\caption{Correspondence between the symmetry data.}
\renewcommand{\arraystretch}{1.8}
\begin{tabular}{m{50pt}<{\centering}m{80pt}<{\centering}m{100pt}<{\centering}m{80pt}<{\centering}m{120pt}<{\centering}m{50pt}<{\centering}}
\hline \hline 
                                 & Symmetry action & Intermediate quantity& Obstruction & Symmetry fractionalization & Torsor \\ \hline 
TR SET                &   $\rho_{\bT} = \bar{\rho}_{\mb}$              &  $[U_{\bT}(a, b; c)]_{\alpha_1 \alpha_2}$                                   & $\mathcal{O}(\bT, \bT, \bT)$                                &      $\eta_a(\bT, \bT)$           & $\mathfrak{w}(\bT, \bT)$      \\ 
Mirror SET & $\rho_{\mb}=\bar{\rho}_{\bT}$                & ${\phi}^{a_A b_A}_{c_A, (\alpha_1, \alpha_2)}$                                    & $\mathcal{O}(\mb, \mb)$                               & $U_{\bM}(a_A)$               & $\mathfrak{v}(\mb)$      \\ \hline \hline
\end{tabular}
\label{tab:data}
\end{table*}
In this section, we present a detailed methodology for calculating the potential $H^2$ obstruction in mirror SET orders. 
Besides, we roughly review time reversal(TR) SET orders by listing a series of consistency equations.
Based on the fact that mirror symmetry is analogous to TR symmetry through Wick rotation in topological field theory, we conduct a comparative analysis of consistency equations between TR SET orders and their mirror counterparts. 
Consequently, we establish precise correspondence between their symmetry data, as outlined in Table~\ref{tab:data}.

For mirror SET orders, the potential $H^2$ obstruction appears along the mirror axis.   
Subsequently, we apply the formulation of a symmetric gapped boundary to calculate this obstruction in bilayer geometry. 
Firstly, we solve the following two consistency equations associated with VLCs:
\begin{widetext} 
\begin{gather}
    \phi^{a_A b_A}_{c_A,(\alpha_1, \alpha_2)}=\sum_{(\beta_1, \beta_2)} \phi^{b_A a_A}_{c_A,(\beta_1, \beta_2)} [R^{ab}_{c}]_{\beta_1 \alpha_1} [R^{{\bha}{\bhb}}_{\bhc}]_{\beta_2 \alpha_2} \label{71}, \\
    \sum_{e,\alpha_1,\alpha_2,\beta_1,\beta_2} \phi^{a_Ab_A}_{e_A,(\alpha_1, \alpha_2) } \phi^{e_A c_A}_{d_A,(\beta_1,\beta_2) }[F^{abc}_d]_{(e,\alpha_1,\beta_1)(f,\mu_1,\nu_1)}[F^{{\bha}{\bhb}{\bhc}}_{\bar{\rho}_{\mb}(d)}]_{({\bar{\rho}_{\mb}(e)},\alpha_2,\beta_2)({\bar{\rho}_{\mb}(f)},\mu_2,\nu_2)}= \phi^{b_A c_A}_{f_A,(\mu_1, \mu_2)} \phi^{a_A f_A}_{d_A,(\nu_1, \nu_2)}	\label{72}.
\end{gather}
\end{widetext}
Here, we designate the fusion channel as a pair $(\alpha_1, \alpha_2)$ for the bilayer system. 
Secondly, we we substitute the calculated VLCs into the subsequent equation to derive $U_{\bM}(a_A)$
\begin{gather}
    \phi^{a_A b_A}_{c_A,(\alpha_1, \alpha_2)} = \phi^{{\bha}_A{\bhb}_A}_{{\bhc}_A; (\alpha_1, \alpha_2)} \frac{U_{\bM}(c_A)}{U_{\bM}(a_A)U_{\bM}(b_A)}  \label{73},
\end{gather}
where we fix $U_{\be}(a_A) = 1$.
Thirdly, we utilize the obtained $U_{\bg}(a_A)$ to calculate $H^2$ obstruction.
The nontrivial quantities are
 \begin{gather}
     O_{a_A}(\bM, \bM)=\frac{1}{U_{\bM}(a_A)U_{\bM}({\ubbm a}_A)}\label{74}. 	
 \end{gather}
Finally, we employ the calculated $O_{a_A}(\bM, \bM)$ to determine $\cO(\mb, \mb)$.
Specifically, if $\cO(\mb, \mb)$ can't be expressed as a coboundary, $\tau(\mb){\bar{\rho}_{\mb}(\tau(\mb))}$, the determined $H^2$ obstruction, $\cO(\mb, \mb)$, is deemed nontrivial.
Otherwise, $\cO(\mb, \mb)$ can be eliminated by gauge transformation where $H^2$ obstruction vanishes. 
 
Moving on to TR SET orders, we recognize that time reversal symmetry, denoted as $\mathbb{Z}^{\bT}_2 = \{\be, \bT\}$, is an anti-unitary symmetry.
Consequently, we define a nontrivial symmetry permutation for anyons of $\cC$ as follows:
\begin{gather}
     \rho_{\bT}(a) = \bar{\rho}_{\mb}(a)	.
\end{gather}
Furthermore, we establish a nontrivial symmetry action on the topological space $V^{a b}_c$, which is defined by
\begin{align}
	\rho_{\bT}\ket{a, b; c, \mu} = \sum_{\nu}[U_{\bT}(^{\bT}a, ^{\bT}b; ^{\bT}c)]_{\mu \nu}\ket{^{\bT}a, ^{\bT}b; ^{\bT}c, \nu}
\end{align}
Here, the bulk $U$-symbols satisfy two subsequent equations,
\begin{widetext}
\begin{align}
	\sum_{e, \alpha_1, \alpha_2, \beta_1, \beta_2}[U_{\bT}(a, b; e)]_{\alpha_1 \alpha_2} [U_{\bT}(e, c; d)]_{\beta_1 \beta_2} & [F^{a b c}_d]_{(e, \alpha_1, \beta_1)(f, \mu_1, \nu_1)} [F^{{^{\bT}a} {^{\bT}b} {^{\bT}c}}_{^{\bT}d}]_{({^{\bT}e}, \alpha_2, \beta_2)({^{\bT}f, \mu_2, \nu_2})}\notag \\
	  &=[U_{\bT}(b, c; f)]_{\mu_1 \mu_2} [U_{\bT}(a, f; d)]_{\nu_1, \nu_2} \label{77},\\
	\sum_{\beta_1, \beta_2} [U_{\bT}(b, a; c)]_{\beta_1 \beta_2} [R^{a b}_c]_{\beta_1 \alpha_1} [R^{{^{\bT}a} {^{\bT}b}}_{^{\bT}c}]_{\beta_2 \alpha_2} & = [U_{\bT}(a, b; c)]_{\alpha_1 \alpha_2} \label{78}. 
	\end{align}
\end{widetext}
Besides, a consistency equation exists between bulk $U$-symbols and $\eta$-symbols.
\begin{gather}
	\frac{\eta_c(\bT, \bT)}{\eta_a(\bT, \bT) \eta_b(\bT, \bT)} [U_{\bT}({^{\bT}a}, {^{\bT}b}; {^{\bT}c})]_{\alpha_1 \alpha_2} = [U_{\bT}(a, b; c)]_{\alpha_1 \alpha_2} \label{79}.
\end{gather}
For the on-site SET order, $H^3$ obstruction is defined by 
\begin{align}
	\Omega_a(\bT, \bT, \bT) = \frac{1}{\eta_a(\bT, \bT) \eta_{^{\bT}a}(\bT, \bT)}\label{80}.
\end{align}
In particular, when the $H^3$ obstruction vanishes, the $\eta$-symbols can be modified by a torsor $\mathfrak{w} \in H^2_{\rho}[\mathbb{Z}^{\bT}_2, \cA_{\cC}] $ which characterizes the symmetry fractionalization class.
We note that \Eqss{77}{80} determine symmetry datas of TR SET orders. 
In contrast to \Eqss{71}{74}, which describe mirror SET orders, we have identified a precise correspondence between the symmetry data presented in Table~\ref{tab:data}.
Specifically, based on the derived relative t'Hooft anomaly indicator for mirror SET orders as discussed in Sec.~\ref{for}, we are able to directly infer the relative t'Hooft anomaly indicator for TR SET orders through the correspondence.
\begin{gather}
    \eta_r = \eta_{\mathfrak{w}(\bT, \bT)}(\bT, \bT)\theta_{\mathfrak{w}(\bT, \bT)},
\end{gather}
 which is exact result in Ref.~\cite{barkeshli2020relative}.

\section{An anomalous example: $\mathbb{D}_{16}$ gauge theory}
\label{sec:d16}
\begin{table*}[ht]
\centering
\caption{Anyons of $\mathbb{D}_{16}$ gauge theory.}
\renewcommand{\arraystretch}{1.5}
\begin{tabular}{m{80pt}<{\centering}m{60pt}<{\centering}m{50pt}<{\centering}m{60pt}<{\centering}m{50pt}<{\centering}m{60pt}<{\centering}m{50pt}<{\centering}m{15pt}<{\centering}}
\hline\hline 
Conjugate class & $[\be]$ & $[\br]$ & $[\ba]$  & $[{\ba}^2]$ & $[{\ba}^4]$ &      $[\br \ba]$  & $[{\ba}^3]$    \\ 
Representation  & $\phi_i, \alpha_j$  & $\psi_i$ & $\beta_i$ & $\beta_i$ & $\phi_i, \alpha_j$ & $\psi_i$ & $\beta_i$     \\ \hline\hline
\end{tabular}
\label{tab:anyon}
\end{table*}
In this section, we investigate a specific anomalous example with nontrivial $H^2$ obstruction.
This example is $\mathbb{D}_{16}$ gauge theory, denoted as $\cC$, which is enriched by mirror symmetry $\mathbb{Z}^{\mb}_2$.
We calculate the $H^2$ obstruction in bilayer geometry and subsequently verify its correctness by correspondence between TR SET orders and mirror counterparts.
Moreover, we derive the child topological order $\cD = \text{Rep}(D(\mathbb{Z}_2))$ which can be utilized to eliminate the $H^2$ obstruction.
 
Following notation in Ref.~\cite{qi2019folding}, the anyons of $\cC$ are labeled by the pair $([\bg], \phi)$.
Here, $[\bg]$ signifies a conjugacy class of the dihedral group $\mathbb{D}_{16}$ with $\bg$ representing an element from this class. 
The $\phi$ denotes an irreducible representation of the centralizer group, defined as $Z_{\bg} = \{\bh|\bh \bg=\bg \bh, \bh\in \mathbb{D}_{16}\}$. 
Additionally, the elements of $\mathbb{D}_{16}$ are represented by the set $\{\br^n \ba^m|\br^2=\ba^8=1, \br \ba \br=\ba^{-1}   \}$. 
There are seven conjugacy classes: $[\be], [\br], [\br \ba], [\ba], [\ba^2], [\ba^3], [\ba^4]$. 
Analyzing irreducible representations of corresponding centralizer groups, respectively, there are 46 anyons. 
According to the data presented in Table~\ref{tab:anyon}, we can label the 46 different anyons.
In this table, $\phi_i$ denote four one-dimensional irreducible representations of the dihedral group $\mathbb{D}_{16}$, and $\alpha_j$ denote three two-dimensional irreducible representations of the same group.
Additionally, $\beta_i$ represent eight one-dimensional representations of the cyclic group $\mathbb{C}_8$, and $\psi_i$ represent four one-dimensional representations of the dihedral group $\mathbb{D}_4$. 
More specifically, those are
\begin{equation*}
\begin{gathered}
     \phi_1(\bg) = 1,~\phi_2(\br) =-1,~\phi_3(\ba)=-1,\notag\\
	 \phi_4(\br)=\phi_4(\ba)=-1,~ \alpha_j(\ba)=\text{diag}[e^{\frac{ij\pi}{4}}, e^{-\frac{ij\pi}{4}}], \notag \\
	 \psi_1(\bg)=1,~\psi_2(\br {\ba}^n)=-1,~\psi_3({\ba}^4)=-1, \notag \\
	 \psi_4({\ba}^4)=\psi_4(\br {\ba}^n)=-1,~\beta_k(\ba)=e^{\frac{ik\pi}{4}	},
\end{gathered}
\end{equation*}
where $j = 1,2,3$ and $k = 0,1,2,3,4,5,6,7$.   

We now consider a group automorphism $f \in \text{Out}(\mathbb{D}_{16})$ defined by the following actions:
\begin{align*}
	f(\ba) = {\ba}^5,~ f(\br) = \br \ba.
\end{align*}
This automorphism induces a nontrivial linear auto-equivalence $\xi^f$ on $\cC$, characterized by the mapping
\begin{align*}
	\xi^f : ([\bg], \phi) \mapsto ([f(\bg)], \phi \circ f^{-1} ).
\end{align*}
Here, $\phi \circ f^{-1}$ represents a representation of the centralizer $Z_{f(\bg)}$, satisfying $\phi \circ f^{-1}(f(\bg))=\phi(\bg)$. 
Subsequently, we define the anti-linear mirror symmetry action as follows:
\begin{align}
	\bar{\rho}_{\mb} : ([\bg], \phi) \mapsto ([f(\bg)], (\phi \circ f^{-1} )^* ).
\end{align}
We present a list of nontrivial permutations,
\begin{equation*}
\begin{gathered}
	\bar{\rho}_{\mb}([\be], \phi_3)=([\be], \phi_4),~ {\bar{\rho}_{\mb}([{\ba}^4], \phi_3)}=([{\ba}^4], \phi_4), \notag \\
	\bar{\rho}_{\mb}([\be], \alpha_1)=([\be], \alpha_3),~ {\bar{\rho}_{\mb}([{\ba}^4], \alpha_1)}=([{\ba}^4], \alpha_3), \notag \\
	\bar{\rho}_{\mb}([\ba], \beta_0)=([{\ba}^3], \beta_0),~ {\bar{\rho}_{\mb}([\ba], \beta_4)=([{\ba}^3], \beta_4)}, \notag \\
	\bar{\rho}_{\mb}([\ba], \beta_1)=([{\ba}^3], \beta_5),~ {\bar{\rho}_{\mb}([\ba], \beta_3)=([{\ba}^3], \beta_7)}, \notag \\
	\bar{\rho}_{\mb}([\ba], \beta_2)=([{\ba}^3], \beta_2),~ \bar{\rho}_{\mb}([\ba], \beta_6)=([{\ba}^3], \beta_6),\notag \\
	\bar{\rho}_{\mb}([\br \ba], \psi_i)=([\br], \psi_i),~ {\bar{\rho}_{\mb} ([{\ba}^2], \beta_1)}=([{\ba}^2], \beta_3), \notag \\
	{\bar{\rho}_{\mb} ([{\ba}^2], \beta_2)}=([{\ba}^2], \beta_6), {\bar{\rho}_{\mb} ([{\ba}^2], \beta_5)}=([{\ba}^2], \beta_7).  
\end{gathered}
\end{equation*}
Because anyons in $\cC$ are self-dual, we derive
\begin{gather*}
     O_{a_A}(\bM, \bM) = U^2_{\bM}(a_A) = 1.	
\end{gather*}
Here, anyon $a$ is invariant under the $\rho_{\mb}$-action.
According to the fusion rules of anyons, boundary $U$-symbols of those eight $\rho_{\mb}$-invariant anyons are interconnected by \Eq{73},
\begin{equation*}
\begin{gathered}
	U_{\bM}(([\be], \phi_1)_A)=U_{\bM}(([\be], \phi_2)_A)=1,\\
    U_{\bM}(([{\ba}^4], \phi_1)_A)=U_{\bM}(([{\ba}^4], \phi_2)_A)=U_{\bM}(([{\ba}^2], \beta_0)_A),\\
    U_{\bM}(([{\ba}^4], \phi_1)_A)=U_{\bM}(([\be], \alpha_2)_A)U_{\bM}(([{\ba}^4], \alpha_2)_A),\\
    U_{\bM}(([{\ba}^2], \beta_4)_A)=U_{\bM}(([{\ba}^4], \alpha_2)_A).
\end{gathered}
\end{equation*}
Thus, the independent boundary $U$-symbols of $\rho_{\mb}$-invariant anyons are $U_{\bM}(([{\ba}^4], \alpha_2)_A)$ and $U_{\bM}(([\be], \alpha_2)_A)$. 
Applying \Eq{73} to the anyons $([\be], \alpha_1)_A$ and $([\be], \alpha_3)_A$, we obtain
\begin{gather}
	O_{([\be], \alpha_1)}(\bM, \bM)=U_{\bM}(([\be], \alpha_2)_A),\\
	U_{\bM}(([\be], \phi_3)_A)=U_{\bM}(([\be], \alpha_2)_A)\frac{\phi^{([\be], \alpha_1)_A([\be], \alpha_3)_A}_{([\be], \phi_3)_A}}{\phi^{([\be], \alpha_3)_A([\be], \alpha_1)_A}_{([\be], \phi_4)_A}} \label{86}.
\end{gather}
For the anyons $([\be], \alpha_2)_A$ and $([\be], \phi_3)_A$, we apply \Eq{73} once more to derive
\begin{gather}
	U_{\bM}(([\be], \phi_3)_A)=\frac{\phi^{([\be], \phi_4)_A([\be], \alpha_2)_A}_{([\be], \alpha_2)_A}}{\phi^{([\be], \phi_3)_A([\be], \alpha_2)_A}_{([\be], \alpha_2)_A}} \label{87}.
\end{gather}
According to \Eq{71}, for pure charges $a, b, c$, we recognize the property $\phi^{a_Ab_A}_{c_A} = \phi^{b_Aa_A}_{c_A} $. 
Subsequently, applying \Eq{72} to pure charges, we obtain
\begin{align}
	{\hat{\phi}}^{\alpha_2 \phi_2}_{\alpha_2}{\hat{\phi}}^{\alpha_2 \phi_3}_{\alpha_2}&[{\hF}^{\alpha_2\phi_2\phi_3}_{\alpha_2}]_{\alpha_2, \phi_4}[{\hF}^{\alpha_2\phi_2\phi_4}_{\alpha_2}]_{\alpha_2, \phi_3} \notag \\ 
	     =&{\hat{\phi}}^{\phi_2\phi_3}_{\phi_4}{\hat{\phi}}^{\alpha_2\phi_4}_{\alpha_2} \label{88}, \\
	{\hat{\phi}}^{\phi_2 \alpha_3}_{\alpha_3}{\hat{\phi}}^{\alpha_3 \alpha_1}_{\phi_4}&[{\hF}^{\phi_2\alpha_3\alpha_1}_{\phi_4}]_{\alpha_3, \phi_3}[{\hF}^{\phi_2\alpha_1\alpha_3}_{\phi_3}]_{\alpha_1, \phi_4} \notag \\ 
	     =&{\hat{\phi}}^{\alpha_3\alpha_1}_{\phi_3}{\hat{\phi}}^{\phi_2\phi_3}_{\phi_4} \label{89}, \\
	{\hat{\phi}}^{\phi_2\alpha_3}_{\alpha_3}{\hat{\phi}}^{\alpha_3 \alpha_3}_{\alpha_2}&[{\hF}^{\phi_2\alpha_3\alpha_3}_{\alpha_2}]_{\alpha_3, \alpha_2}[{\hF}^{\phi_2\alpha_1\alpha_1}_{\alpha_2}]_{\alpha_1, \alpha_2} \notag \\
	    =&{\hat{\phi}}^{\alpha_3\alpha_3}_{\alpha_2}{\hat{\phi}}^{\phi_2\alpha_2}_{\alpha_2} \label{90}.     
\end{align} 
For notational convenience, we denote the $R$-symbols, $F$-symbols and VLCs associated with pure charges as $\hR, \hF$, and $\hat{\phi}$, respectively. 
Based on \Eqss{86}{90}, we derive
\begin{align}
	U_{\bM}(([\be], \alpha_2)_A) = -1.
\end{align}
According to the cocycle condition of the $H^2$ obstruction $\cO(\mb, \mb)={\ubm \cO(\mb, \mb)}$, we conclude
\begin{gather}
 	\cO(\mb, \mb)=([{\ba}^4], \phi_1) \label{92}.
\end{gather}
We note that there exists an alternative result that differs by a coboundary when compared to $([{\ba}^4], \phi_1)$.
Thus, the two results are equivalent. 

According to the result in Ref.~\cite{fidkowski2017realizing}, a nontrivial $H^3$ obstruction, characterized by $([{\ba}^4], \phi_1)$, is present when $\text{Rep}(D(\mathbb{D}_{16}))$ is enriched with an on-site $\mathbb{Z}_2$ symmetry where non-trivial permutation is defined by $\xi^f$.
Then, we examine the associated TR SET orders where the non-trivial permutation $\rho_{T}$ is specified by the composition of $\xi^f$ and non-anomalous permutation $\bar{\rho}_1$, defined as $\bar{\rho}_1 : ([\bg], \phi) \rightarrow ([\bg], (\phi)^* )$~\cite{qi2019folding}.
It implies that the $H^3$ obstruction is given by $\cO(\bT, \bT, \bT)=([{\ba}^4], \phi_1)$.
Since $\rho_{\bT} = \bar{\rho}_{\mb}$, we directly obtain $H^2$ obstruction of the associate mirror SET order through correspondence detailed in Table~\ref{tab:data},
\begin{align}
	\cO(\mb, \mb)= \cO(\bT, \bT, \bT)= ([{\ba}^4], \phi_1),
\end{align}
which is the exact result of \Eq{92}.

Now, we proceed to eliminate the $H^2$ obstruction by employing the approach discussed in Sec.~\ref{fif}.
The condensable subalgebra $B = \oplus_{(b, \bar{\rho}_{\mb}(b))} (b, \bar{\rho}_{\mb}(b))$ is determined by condition $M_{b\cO(\mb, \mb)} = 1$,
\begin{align}
	M_{([\bg], \nu)([{\ba}^4], \phi_1)} = \frac{\text{tr }\nu({\ba}^4)}{\text{tr }\nu(\be)} \label{94},
\end{align}
where we refer to the S-matrix in Ref.~\cite{bakalov2001lectures}.
Thus, the dimension of $B$ is derived  
\begin{align}
	\text{FPdim}(B)= 128.
\end{align}
Moreover, the total dimension of child topological order is calculated as $\text{Dim}(\cD)=\frac{\text{Dim}(\cC \boxtimes \cC)}{\text{FPdim}(B)^2}=4$. 
Besides, anyons $(a, \bar{\rho}_{\mb}(a))$, satisfying $M_{a\cO(\mb, \mb)}\neq1 $ are mapped to $\mathbb{Z}_2$ pure flux $m_{\cD}$ in $\cD$. 
Hence, we claim 
\begin{align}
	\mathcal{D} = \text{Rep}(D(\mathbb{Z}_2)),
\end{align}
which is in agreement with \Eq{56}.

Subsequently, we introduce the $\mathbb{Z}^{\bM}_2$ symmetry into $\cD$.
This introduction may lead to a nontrivial $H^2$ obstruction appearing on a gapped boundary of $\cD$, which is characterized by the Lagrangian algebra $\tilde{A} = 1 \oplus m_{\mathcal{D}}$.
Specifically, we have
\begin{align}
	o_{m_{\cD}}(\bM, \bM) = \eta^{-1}_{m_{\cD}}(\bM, \bM) = M_{m_{\cD}\mathfrak{w}(\bM, \bM)}.
\end{align}
Here, $\mathfrak{w}$ is a torsor which can modify $\mathbb{Z}^{\bM}_2$ symmetry fractionalization class in the bulk of the child SET order. 
By adjusting $o_{m_{\cD}}(\bM, \bM)$ through the torsor $\mathfrak{w}$, we can match it with the $H^2$ obstruction on the boundary of the parent SET order. 
It follows that 
\begin{align}
	\mathfrak{w}(\bM, \bM) = (([{\ba}^4], \phi_1), 1_{\cC}).
\end{align}
Hence, the two anomalous boundaries can be merged to form an interface that makes the $H^2$ obstruction vanishing. 

\section{Conclusion}
\label{sec:conclusion}

In this work, we present a concrete description of mirror-SETs using the folding approach.
In particular, we give consistency conditions among data describing the mirror-symmetry action on the mirror axis, which encodes the classification of mirror-SET states in the folding appraoch,
and we derive the formula of an $H^2$ obstruction function from these consistency conditions.
We show that this $H^2$ obstruction function is the counterpart of an $H^3$ obstruction function in time-reversal SET states, whose classification is isomorphic to that of mirror-SETs.
Using a layered construction, we demonstrate that mirror SET states with a nontrivial $H^2$ obstruction can be realized on the surface of a 3D mirror-SET state, which resembles the boundary-bulk relation between onsite-SET states with an $H^3$ obstruction and 3D SET states.
This $H^2$ obstruction function, together with the $H^3$ obstruction that is reflected in the topological spin of the $\mathbb Z_2$ symmetry defects, completely characterizes the anomalies of the 2D mirror-SET states.

It will be interesting to study how the $H^2$ obstruction can be generalized to fermionic mirror-SETs, and to SETs protected by a combination of mirror and other onsite symmetries.
Moreover, it is desirable to have a practical algorithm to systematically search for solutions of consistency equations(\ref{71}--\ref{74}).
We will leave these to future works.

\begin{acknowledgments}
	The authors thank Chenjie Wang and Meng Cheng for insightful discussions.
	Y.Q. acknowledges support from National Key R\&D Program of China (Grant No. 2022YFA1403402), from the National Natural Science Foundation of China (Grant No. 12174068), and from the Science and Technology Commission of Shanghai Municipality (Grant No. 23JC1400600).
\end{acknowledgments}

\bibliography{reference}
\end{document}